\documentclass[letterpaper, 10 pt, conference]{ieeeconf}  

\IEEEoverridecommandlockouts                              
\title{\LARGE \bf
 Dynamic Mode Decomposition with Non-uniform Sampling
}

\author{Ramachandran Ananthraman and Alexandre Mauroy
\thanks{Both the authors are with the Department of Mathematics and Namur Institute of Complex Systems (NaXys), University of Namur, 5000 Namur, Belgium. 
{\tt\small ramachandran.chittur@unamur.be and alexandre.mauroy@unamur.be}}%
\thanks{R Anantharaman is a Postdoctoral Researcher at UNamur. This work was supported by the F.R.S.-FNRS grant PDR-T.0148.22}
}

\usepackage{color}
\usepackage{float}
\usepackage{amsmath}
\usepackage{amssymb}
\usepackage{esint}
\usepackage{hyperref}
\usepackage{varioref}
\usepackage{array}
\usepackage{mathtools}

\newtheorem{remark}{Remark}
\newtheorem{assumption}{Assumption}
\usepackage{slashbox}
\usepackage{algorithm}
\usepackage{algorithmic}

\def\Rr{\mathbb{R}}

\def\mcf{\mathcal{F}}
\def\mcb{\mathcal{B}}

\def\mphi{\boldsymbol{\Phi}}
\def\Koop{\mathcal{K}}

\def\bUN{\mathbf{K}_N}
\def\bU{\mathbf{K}}
\def\bL{\mathbf{L}}

\newcommand\norm[1]{\left\lVert#1\right\rVert}

\begin{document}

\maketitle
\thispagestyle{empty}
\pagestyle{empty}
\begin{abstract}
Dynamic Mode Decomposition (DMD) and its extensions (EDMD) have been at the forefront of data-based approaches to Koopman operators. Most (E)DMD algorithms assume that the entire state is sampled at a uniform sampling rate. In this paper, we provide an algorithm where the entire state is not uniformly sampled, with individual components of the states measured at individual (but known) sampling rates. We propose a two-step DMD algorithm where the first step performs Hankel DMD on individual state components to estimate them at specified time instants. With the entire state reconstructed at the same time instants, we compute the (E)DMD for the system with the estimated data in the second step. 

\end{abstract}
\section{Introduction}
Koopman operator \cite{Koopman1} based approaches to nonlinear systems have recently gained significant interest in dynamical systems theory. The advantage of transforming a nonlinear system to a linear one together with its adaptability to data-driven techniques make the Koopman operator theory a promising framework to study nonlinear systems. Dynamic Mode Decomposition (DMD) and their extensions lie in the forefront of most data-based frameworks for the Koopman operator. Initial approaches to DMD \cite{PJS2010,CWR2009} used snapshots of the state vector (the entire state) from a single trajectory at uniform intervals. Later, the work of \cite{JHT2014} extended this paradigm to work with data from multiple data paths as well approximate the Koopman operator over a space spanned by nonlinear observables \cite{MOW2015}. However the assumption that the entire state is measured with a constant sampling period for all the data paths still remained. DMD approaches with non-uniform sampling of data have been discussed in the works of \cite{LC2016, GMP2015} where the entire state data is not available uniformly. The former work addressed the case of missing measurements in uniformly sampled data and estimated the missing data through Bayesian inference, while the latter addressed the setting of non-uniformly sampled data by changing the Vandermonde matrix used in the DMD algorithm to a more general matrix. However, both these approaches still assume that when data is available at a specific time instant, the entire state of the system is available. In this work, we depart from that assumption and suppose that the data available at any time instant contains only partial state measurements. This situation can arise when measurement of the entire state is impossible or expensive, limiting the data available for DMD computation. In this manuscript, we will provide an algorithm for this partial measurement setting which is primarily based on Hankel DMD \cite{SM2015,AM2017}. The algorithm works in two steps. The first step estimates the entire state from the partial measurements by performing Hankel DMD on each state component. Once the entire state is estimated at two different time instants differing by a constant time $T_s$, we perform EDMD with this reconstructed state data in the second step. Further, we illustrate two different applications of this algorithm, defined as (i) Multirate EDMD where each state component is measured at an individual but known sampling frequency and (ii) Single state EDMD where only one component of the state is measured at each sampling instant.


The paper is organized as follows. The following section introduces the preliminaries about the Koopman operator and its data-driven approach. Section \ref{s3} focuses on the EDMD algorithm with non-uniform sampling. Section \ref{s4} presents the two specific applications of the proposed algorithm with numerical simulations on a Lorenz system.  
\section{Preliminaries}
Consider a dynamical system
\begin{align}
    \label{eq:DS}
    \dot{x}(t) = F(x(t)),
\end{align}
where $x \in \Rr^n$ and $F : \Rr^n\to \Rr^n$ are the state and the vector field, respectively. The solution $x(t)$ at time $t$ from an initial condition $x_0$ is given through the flow $S^t$ as 
\[
x(t) = S^t(x_0).
\]
Consider a space of functions $\mcf: \Rr^n \to \Rr$. The Koopman operator $\Koop^t$ \cite{Koopman1} associated with the system (\ref{eq:DS}) is defined as follows
\begin{align}
    \label{eq:Koop}
    \Koop^t h(x) = h(S^t(x)),
\end{align}
for $h \in \mcf$. The Koopman operator is linear over $\mcf$ and provides an interesting viewpoint to study (\ref{eq:DS}) as $\Koop^t$ is linear over $\mcf$ irrespective of the nonlinearity of the dynamics (\ref{eq:DS}). This linearity is traded with the dimensionality, and in general, for dynamics over $\Rr^n$, this space $\mcf$ is infinite-dimensional. Under additional assumptions on the dynamics, the Koopman operator $\Koop^t$ is also a ${C}_0$-semigroup in the space $\mcf$ \cite{AM-IM-YS2020} and allows an infinitesimal generator $L$ to be defined as
\[
L h := \lim_{t \downarrow 0} \frac{\Koop^t h - h}{t},
\]
for all $h$ in the domain of $L$.
\subsection{Data-based approaches to Koopman operator}
In many engineering applications of the Koopman operator, it is necessary to restrict the Koopman operator to a finite-dimensional subspace of $\mcf$ defined through a choice of lifting functions and use this finite-dimensional linear system for analysis and control. This restriction of the Koopman operator to a finite-dimensional subspace can be achieved in a data-driven setting. DMD and its extensions lie in the forefront of such data-based approximations of the Koopman operator, and the original algorithm was initially proposed in \cite{PJS2010} to study nonlinear fluid flows. Its connection to the Koopman operator \cite{JHT2014,CWR2009} and the corresponding applications to data-driven approximations of the Koopman operator lead to the formulation of finite-dimensional linear lifted systems for nonlinear dynamics through the Koopman operator. 

The core idea of DMD-based approaches is given as follows: Consider a subspace $\mcf_N \subset \mcf$ defined by a chosen set of basis functions $\mcb = \{\phi_1,\dots,\phi_N\}$. We proceed to compute $\Koop_N^t : \mcf_N \to \mcf_N$
\begin{align}
\label{eq:KoopRest}
\Koop_N^t := \Pi_N \Koop^t|_{\mcf_N},
\end{align}
where $\Pi_N$ is an projection (most common algorithms use $\mathcal{L}_2$ projections). In particular, we aim to find for $g\in \mcf_N$
\begin{align}
\label{eq:Projection}
\Pi_N \Koop^t g = \min_{\tilde{g} \in \mcf_N} \norm{\Koop^t g - \tilde{g}}.
\end{align}
Apart from the numerical implementations, the various DMD algorithms primarily differ in their choice of lifting functions. For DMD, the basis functions $\phi_i$ are the $n$ identity functions; for other algorithms, suitable choices of functions are used. 

\subsection{Implementation of DMD and computing lifted dynamics}
The initial approaches to DMD used the data that is generated by sampling the solution trajectory $x(t)$ of the system at time $0,T_s,2T_s,\dots,mT_s$ as $x(0),x(1)$, $x(2),\dots,x(K)$, and we have
\begin{align}
\label{eq:DMD_data_OT}
x(k+1) = S^{T_s}(x(k)).
\end{align}
When the projection is $\mathcal{L}^2$ (we assume that all projections in this paper are $\mathcal{L}^2$ unless explicitly mentioned), we can compute the matrix representation $\bUN$ of $\Koop_N^{T_s}$ over $\mcf_N$ with the basis functions $\{\phi_i\}$ using linear least squares formulation. Define
\begin{align}
\label{eq:Mphi}
\mphi = [\phi_1,\dots,\phi_N]^T,
\end{align}
and let
\begin{align}
\label{eq:MatDMD}
\begin{aligned}
\mathbf{P}_x &= \begin{bmatrix} \mphi(x(0))& \dots & \mphi(x(K-1))  \end{bmatrix}\\ 
\mathbf{P}_y &= \begin{bmatrix} \mphi(x(1)) & \dots & \mphi(x({K})) \end{bmatrix},
\end{aligned}
\end{align}
and we have
\[
\mathbf{P}_y = \bUN \mathbf{P}_x. 
\]
When $\mathbf{P}_x$ is full rank ($K \gg N$), we compute $\bUN$ as 
\begin{align}
\label{eq:KoopFinite}
\bUN = \mathbf{P}_y \mathbf{P}_x^\dagger.
\end{align}
In the subspace $\mcf_N$, the action of the Koopman operator $\Koop^{T_s}$ is approximated through the matrix $\bUN$ as 
\begin{align}
\label{eq:KoopBasis}
\begin{bmatrix}
    \Koop^{T_s} \phi_1 \\ \vdots \\  \Koop^{T_s} \phi_N 
\end{bmatrix} \approx \bUN \begin{bmatrix}
 \phi_1 \\ \vdots \\   \phi_N 
\end{bmatrix}.
\end{align}
Also, the infinitesimal generator $L$ can be approximated as $\bL_N$ over $\mcf_N$, where $\bL_N$ is defined as \cite{MG2020}
\begin{align}
\label{eq:InfGen}
\bL_N = \frac{1}{T_s}log(\bUN).
\end{align}
With this infinitesimal generator, we can compute the time-derivatives of the basis functions as 
\[
\begin{bmatrix}
    \dot{\phi}_1\\ \vdots \\  \dot{\phi}_N
\end{bmatrix} \approx \bL_N \begin{bmatrix}
 \phi_1\\ \vdots \\   \phi_N
\end{bmatrix}.
\]
Let $\bUN^t$ be the matrix representation of the $\Koop_N^t$ as in (\ref{eq:KoopRest}) at any time $t$ with the basis functions $\{\phi_i\}$. Using this $\bL_N$, one can compute the approximation of the $\bUN^t$ as
\begin{align}
\label{eq:Koop_InfG}
\bUN^t = e^{\bL_N t}.
\end{align}
It is to be noted that the matrices $\bL_i$ and $\bUN^t$ defined through equations (\ref{eq:InfGen}) and (\ref{eq:Koop_InfG}) are the approximations of the true matrix representations corresponding to the respective projections of the operators $\Pi_N L|_{\mcf_N}$ and $\Pi_N \Koop^t_{\mcf_N}$ \cite{MG2020}. In general DMD algorithms \cite{MOW2015}, the data can be collected over multiple trajectories, and we can assume the data is available as $\{x^{(k)},y^{(k)}\}_{k=1}^K$ such that $y^{(k)}$ satisfies 
\begin{align}
\label{eq:data_DMD}
y^{(k)} = S^{T_s}(x^{(k)}).
\end{align}
Given the basis functions $\{\phi_i\}$, for the data $\{x^{(k)},y^{(k)}\}_{k=1}^K$, the matrices $\mathbf{P}_x$ and $\mathbf{P}_y$ defined in (\ref{eq:MatDMD}) is computed as 
\begin{align}
\label{eq:matEDMD}
\begin{aligned}
\mathbf{P}_x &= \begin{bmatrix} \mphi(x^{(1)})& \dots & \mphi(x^{(K)})  \end{bmatrix}\\ 
\mathbf{P}_y &= \begin{bmatrix} \mphi(y^{(1)}) & \dots & \mphi(y^{(K)}) \end{bmatrix}.
\end{aligned}
\end{align}

\subsection{Hankel DMD}
Among the variants of EDMD, Hankel DMD \cite{AM2017,SM2015} is of particular interest in this work, and we summarize the method in this section. Given a function $g \in \mcf$, we know that the action of $\Koop^t$ on $g$ is a new function $\Koop^t g \in \mcf$. Given a flow $S^t(x)$, the value of $\Koop^t g$ at finite time instants $kT_s$, $k = 0,\dots,M-1$ can be used as the basis functions for the finite-dimensional space $\mcf_N$. This formulation is useful when the dynamics is associated with an output map $g$ as follows
\begin{align}
\label{eq:DSop}
\begin{aligned}
\dot{x}(t) = F(x(t)) \\ 
z(t) = g(x(t)),
\end{aligned}
\end{align}
where $x \in \Rr^n, z \in \Rr$ are the states and output of the dynamics, $F$ and $g$ are the vector field and the output map respectively. Choosing the lifting functions to be 
\begin{align}
\label{eq:BasisHDMD}
\phi_1 = g, \phi_2 =  \Koop^{T_s} g, \dots, \phi_M = \Koop^{(M-1)T_s}g,
\end{align} 
it can be seen that 
\[
\Koop^{T_s} \begin{bmatrix} g \\   \vdots \\ \Koop^{(M-1)T_s}g\end{bmatrix} = \begin{bmatrix} \Koop^{T_s} g \\   \vdots \\ \Koop^{MT_s}g\end{bmatrix}.
\]
Given $M$ measurements $z(0),z(T_s),\dots,z((M-1)T_s)$ of the output over a trajectory of the system (\ref{eq:DSop}), we can evaluate the functions in equation (\ref{eq:BasisHDMD}) as 
\[
\Koop^{kT_s}g(x) = z(kT_s).
\]
Define
\[
\mphi = [g, \Koop^{T_s} g, \dots , \Koop^{(M-1)T_s}g]^T,
\]
and 
\[
\Koop^{T_s}\mphi = [\Koop^{T_s}g, \Koop^{2T_s} g, \dots , \Koop^{MT_s}g]^T.
\]
From $K$ such trajectories and the measurement of the corresponding outputs for $M+1$ time instants $z^{(k)}(0),z^{(k)}(T_s),$ $\dots,z^{(k)}((M-1)T_s),z^{(k)}(MT_s)$, we can construct the matrices $\mathbf{P}_x$ and $\mathbf{P}_y$ as follows:
\begin{align}
\begin{aligned}
\mathbf{P}_x &= \begin{bmatrix} \mphi(x^{(1)}) & \dots & \mphi(x^{(K)} \end{bmatrix} \\
\mathbf{P}_y &= \begin{bmatrix} \Koop^{T_s}\mphi(x^{(1)}) & \dots & \Koop^{T_s}\mphi(x^{(K)} \end{bmatrix}.
\end{aligned}
\end{align}
With these matrices, using equation (\ref{eq:KoopFinite}), we can construct the matrix representation of $\Koop^{T_s}$ on the space spanned by the functions (\ref{eq:BasisHDMD}) as $\bUN$.

\section{EDMD with non-uniform sampling of states}
\label{s3}
During the computation of EDMD, the standard assumption is that the data (\ref{eq:data_DMD}) is available at a fixed sampling rate $T_s$, because the matrix $\bUN$ computed in (\ref{eq:KoopFinite}) is an approximation of $\Koop^{T_s}$ over the subspace $\mcf_N$. In practical settings, this can be restrictive as the measurement of certain components of the state $x$ could be more expensive or time-consuming than the other components, leading to non-uniform availability of samples. To relax this setting, we assume that each component $x_i$ of the state $x$ can be measured at an individual sampling period $T_i$ with multiple samples along a trajectory and the can be experiment repeated over multiple trajectories $x^{(k)}$. We state the setting for performing EDMD with the following assumption. 
\begin{assumption}
\label{A1}
In each trajectory $x^{(k)}$, the data for each state component $x_i^{(k)}$ is available at time instants $r_i+lT_i$, with $l = 0,1,\dots,M_i$, $r_i \in \Rr_+$ and $T_i \in \Rr_+$ are the dead time in measurement of the first sample and sampling period respectively. $M_i+1$ is the number of samples of the component $i$ available over each trajectory $k$. The total number of trajectories is assumed to be $K$. 
\end{assumption}

In this setting, we will devise an algorithm to perform EDMD for the system and approximate $U^{T_s}$ for some arbitrary time $T_s$, and construct the lifted linear system as in (\ref{eq:KoopBasis}). The algorithm consists of the following two steps:
\begin{enumerate}
\item In the first step, for a given $T_s$, we estimate each component $x_i$ of the $k^{th}$ sample trajectory $x^{(k)}$ at time instants $T_s$ and $2T_s$ independently through Hankel DMD. 
\item Once the estimates of $x^{(k)}$ are computed at time $T_s$ and $2T_s$, we have 
\[
x^{(k)}(2T_s) = S^{T_s}{x^{(k)}(T_s)} \quad \forall k = 1,\dots,K.
\]
Defining $y^{(k)} = x^{(k)}(2T_s)$, we have the data as $\{x^{(k)},y^{(k)}\}_{k=1}^K$. Choosing a finite set of lifting functions $\{\phi_i\}_{i=1}^N$, $\phi_i \in \mcf$, we can perform EDMD with matrices defined as in (\ref{eq:matEDMD}). 
\end{enumerate}
In this section, we provide the framework to achieve the first step. Once the state estimates are estimated, the second amounts at performing EDMD with the estimated data, which is well studied in the literature. 
\subsection{Estimate of the states through Hankel DMD}
Let $\chi_i \in \mcf$ be the $i^{th}$ identity function (i.e, $\chi_i(x) = x_i$). For each trajectory, we have $M_i+1$ measurements, and we consider the basis functions 
\[
\mcb_i = \{\chi_i, \Koop^{T_i}\chi_i, \dots, \Koop^{(M_i-1)T_i}\chi_i\},
\]
which span the subspace
\[
\mcf_i = \mbox{span} \{\mcb_i\}.
\]
We define $\Koop_i^{T_i}$ as
\[
\Koop_i^{T_i} = \Pi_i \Koop^{T_i}|_{\mcf_i},
\]
where $\Pi_i: \mcf \to \mcf_i$ is the projection defined in (\ref{eq:Projection}). We aim to compute the matrix representation of $\Koop_i^{T_i}$, denoted as $\bU_i$ with the data.

\subsection*{Computation of Hankel DMD and $\bU_i$}
Given the basis functions $\mcb_i$, we define
\[
\mphi_i = [\chi_i,\dots,\Koop^{(M_i-1)T_i}\chi_i]^T.
\]
Given the measurements of $x_i^{(k)}$ along the sample path $k$ at time instants $r_i+lT_i$, we have the evaluation of $\mphi_i$ as 
\small{\begin{align}
\label{eq:BasisHDMDx_i}
\mphi_i(x_i^{(k)}(r_i)) = [x_i^{(k)}(r_i),\dots,x_i^{(k)}(r_i+(M_i-1)T_i)]^T.
\end{align}}
\normalsize{}
Similarly, we can compute the evaluation of $\Koop^{T_i}\mphi_i $ over the sample path $k$ as 
\small{\[
\Koop^{T_i}\mphi_i(x_i^{(k)}(r_i))= [x_i^{(k)}(r_i+T_i),\dots,x_i^{(k)}(r_i+M_iT_i)]^T. 
\]}
\normalsize{}
This allows us to construct the data matrices
\begin{align}
\label{eq:matHDMDxi}
\begin{aligned}
\mathbf{P}_{x_i} &= \begin{bmatrix} \mphi_i(x_i^{(1)}(r_i))& \dots & \mphi_i(x_i^{(K)}(r_i))  \end{bmatrix}\\ 
\mathbf{P}_{y_i} &= \begin{bmatrix} \Koop^{T_i}\mphi_i(x_i^{(1)}(r_i)) & \dots & \Koop^{T_i} \mphi_i(x_i^{(K)}(r_i)). \end{bmatrix}
\end{aligned}
\end{align}
Using these matrices (\ref{eq:matHDMDxi}), from equation (\ref{eq:KoopFinite}), $\bU_i$ can be computed as 
\begin{align}
\label{eq:matU_i}
\bU_i = \mathbf{P}_{y_i} \mathbf{P}_{x_i}^{\dagger}.
\end{align}
From $\bU_i$, and the knowledge of $T_i$, we can compute the approximation of the infinitesimal generator $\bL_i$ on the space $\mcf_i$ using equation (\ref{eq:InfGen}) as 
\begin{align}
\label{eq:matL_i}
\bL_i = \frac{1}{T_i}log(\bU_i).
\end{align}
With this $\bL_i$, from equation (\ref{eq:Koop_InfG}) we can compute the matrix approximation of the Koopman operator $\Koop^{t}$ for any time $t$ on the space $\mcf_i$ as 
\begin{align}
\label{eq:Koopt_i}
\bU_i^{t} = e^{\bL_i t}.
\end{align}
\vspace{-.2in}
\begin{remark}
An alternative way to compute the matrix approximation $\bU_i^{t}$ would be to compute the rational power of the matrix $\bU_i$ defined in (\ref{eq:matU_i}) as 
\[
\bU_i^{t} = (\bU_i)^{\frac{t}{T_i}}.
\]
This method would in principle eliminate the errors associated with the computation of matrix logarithm but care should be taken that the resulting matrix $\bU_i^t$ is real valued as a complex valued $\bU_i$ will lead to a complex valued state estimates computed through (\ref{eq:x_ipred}). In numerical simulations, it was observed that the rational power of $\bU_i$ was complex valued in most cases, making it incompatible in the further computations. 
\end{remark}

\subsection*{Estimation of $x_i$ at time $t = T_s$ and $2T_s$}
In this section, we use equation (\ref{eq:Koopt_i}) to estimate $x_i(t)$ at time $T_s$ and $2T_s$ from the measurements of $x_i$ available as in Assumption 1. For any time $t$, from equation (\ref{eq:KoopBasis}) we can write
\small{\begin{align*}
\begin{bmatrix} x_i(t+r_i) \\ \vdots \\ x_i((M_i-1)T_i+t+r_i) 
\end{bmatrix} 
 &\approx \bU_i^t \begin{bmatrix}x_i(r_i) \\ \vdots \\ \Koop^{(M_i-1)T_i}x_i(r_i) \end{bmatrix} \\
&= \bU_i^t \mphi_i(x_i(r_i)),
\end{align*}}
\normalsize{}
\noindent where $\mphi_i(x_i(r_i))$ is defined as in (\ref{eq:BasisHDMDx_i}). From the above equation, we can see that $\bU_i^t \mphi_i(x_i(r_i))$ estimates $\mphi_i(x_i)$ after a time $t$ from the measurements $\mphi_i(x_i(r_i))$. 
Using $t = T_s - r_i$ in equation (\ref{eq:Koopt_i}), we have the estimate of $\bU_i^{T_s-r_i}$ and compute $\mphi_i(x_i(T_s))$ as
\begin{align*}
\small{
\begin{aligned}
    \mphi_i(\hat{x}_i(T_s)) = \begin{bmatrix}\hat{x}_i(T_s) \\ \vdots \\ \hat{x}_i((M_i-1)T_i+T_s) \end{bmatrix} = \bU_i^{T_s-r_i}  \mphi_i(x_i(r_i)).
\end{aligned}}
\end{align*}
\normalsize{}
Similarly, using $t = 2T_s-r_i$, we can compute the estimate of $\mphi_i(\hat{x}_i(2T_s))$. We see that $\hat{x}_i^{(k)}(T_s)$ and $\hat{x}_i^{(k)}(2T_s)$ are the first components of $\mphi_i(\hat{x}_i(T_s))$ and $\mphi_i(\hat{x}_i(2T_s))$ respectively and their estimates at various sample paths $k$ are given as
\begin{align}
\label{eq:x_ipred}
\small{\begin{aligned}
\begin{bmatrix} \hat{x}_i^{(1)}(T_s) & \dots & \hat{x}_i^{(K)}(T_s) \end{bmatrix} &= e_1^T \bU_i^{T_s - r_i} \mathbf{P}_{x_i} \\
\begin{bmatrix} \hat{x}_i^{(1)}(2T_s) & \dots & \hat{x}_i^{(K)}(2T_s) \end{bmatrix} &= e_1^T \bU_i^{2T_s - r_i} \mathbf{P}_{x_i},
\end{aligned}}
\end{align}
\normalsize{}
where $e_1 = [1,0,\dots,0]^T$ and $P_{x_i}$ is defined as in (\ref{eq:matHDMDxi}). The algorithm to estimate $\hat{x}_i$ at various sample paths is summarized below.

\begin{algorithm}[H]
\caption{Hankel DMD based estimation of $x_i(T_s)$ and $x_i(2T_s)$}
\begin{algorithmic}[1] 
 \renewcommand{\algorithmicrequire}{\textbf{Input:}}
 \renewcommand{\algorithmicensure}{\textbf{Output:}}
 \REQUIRE Data samples $x_i^{(k)}(r_i+lT_i)$, number of samples $M_i$, dead time $r_i$, sampling time $T_i$ and a time $T_s$
 \ENSURE The estimates $\hat{x}_i^{(k)}(T_s)$, $\hat{x}_i^{(k)}(2T_s)$ of $x_i$ at the ${k}^{th}$ sample path at time $T_s$ and $2T_s$.

%

\STATE Given $M_i$, choose the lifting functions as
\[
\chi_i, \Koop^{T_i} \chi_i, \dots, \Koop^{(M_i-1)T_i}\chi_i.
\]

\STATE Define $\mphi_i$ as in equation (\ref{eq:BasisHDMDx_i}) and construct matrices $\mathbf{P}_{x_i}$, $\mathbf{P}_{y_i}$ as in (\ref{eq:matHDMDxi}).
\STATE Compute the matrices $\bU_i$ and $\bL_i$ as in (\ref{eq:matU_i}) and (\ref{eq:matL_i}) respectively. 
\STATE Computing $\bU_i^{T_s-r_i}$ and $\bU_i^{2T_s-r_i}$ from $\bL_i$ as in (\ref{eq:Koopt_i}), the estimates of $\hat{x}_i^{(k)}(T_s)$, $\hat{x}_i^{(k)}(2T_s)$ are computed as in equation (\ref{eq:x_ipred}).
\end{algorithmic}
\label{alg1}
 \end{algorithm}
\subsection{EDMD with estimated data}
With the estimates of $\hat{x}_i^{(k)}(T_s),\hat{x}_i^{(k)}(2T_s)$ for all $i = 1,\dots,n$, we can reconstruct the estimate of the state entire state $\hat{x}^{(k)}$ at time $T_s$ and $2T_s$ such that
\[
\hat{x}^{(k)}(2T_s) \approx S^{T_s}(\hat{x}^{(k)}(T_s)).
\]
Defining 
\[
x^{(k)} = \hat{x}^{(k)}(T_s) \quad \quad y^{(k)} = \hat{x}^{(k)}(2T_s),
\]
we have $K$ pairs of data $\{\hat{x}^{(k)},\hat{y}^{(k)}\}_{k=1}^{K}$ as in (\ref{eq:data_DMD}) to perform the standard EDMD. Choosing a basis functions of the space $\mcf_N$ as $\phi_1,\dots,\phi_N$, we can construct the matrices $\mathbf{P}_x$ and $\mathbf{P}_y$ as in (\ref{eq:matEDMD}). Then, the restriction of the Koopman operator $\bU_N$ on the space $\mcf_N$ can be computed as in (\ref{eq:KoopFinite}).

\section{Applications of EDMD with non-uniform sampling}
\label{s4}
In this section, we propose two specific cases of the non-uniform data availability in practical scenarios and discuss the applicability of the proposed framework in these cases, namely the \emph{multirate sampling} and \emph{single state sampling}. We also illustrate the proposed algorithm with the help of a Lorenz system. 
\subsection{Multirate Sampling}
In the first case, we assume that the data of each state is available at sampling rates $p_iT_s$, which are integral multiples of a base sampling rate $T_s$ and $r_i = 0\ \forall \ i$. Such a situation can occur when some components of the state are measured at a slower rate than others. We call this framework as \emph{multirate EDMD}. In this case, since each of the sampling time $T_i$ are integral multiples of $T$, we have the measurement of the entire state $x$ at $MT_s$, where 
\begin{align}
\label{eq:LCM}
M = lcm(p_1,\dots,p_N).
\end{align}
provided $M_ip_i > M$ for all $i$. An example of such a scenario for $n=3$ is given in Table \ref{tab1} where $p_1 = 1, p_2 = 2$ and $p_3 = 3$. 
\begin{table}
       \centering
     \scalebox{0.65}{ \setlength{\extrarowheight}{8pt}%
     \begin{tabular}{||c|c|c|c|c|c|c|c||}
    \hline
         \backslashbox{States}{Time}&  $0$ & $T_s$ & $2T_s$ & $3T_s$ & $4T_s$ & $5T_s$ & $6T_s$  \\ \hline 
         $x_1$  & \centering{$x_1(0)$} & $x_1(T_s)$ & $x_1(2T_s)$ & $x_1(3T_s)$ & $x_1(4T_s)$ & $x_1(5T_s)$ & $x_1(6T_s)$  \\ \hline
         $x_2$  & $x_2(0)$ & \textcolor{blue}{$\hat{x}_2(T_s)$} & $x_2(2T_s)$ & - & $x_2(4T_s)$ & - & $x_2(6T_s)$  \\ \hline
         $x_3$  & $x_3(0)$ &\textcolor{blue}{$\hat{x}_3(T_s)$} & - & $x_3(3T_s)$ & -  & - & $x_3(6T_s)$  \\ \hline
    \end{tabular}}
    \caption{Multirate sampling of data: The components in black are measured, and those in blue need to be estimated}
    \label{tab1}
   \end{table} 

In this framework, the full state is measured at $x^{(k)}(0)$ and $x^{(k)}(MT_s)$, and a naive idea would be to directly compute $\bU_N^{MT_s}$. However, all the measurements would not be used and there can be a large error in the DMD computation, especially when the underlying dynamics is fast compared to the sampling rate $MT_s$. Hence, it is preferable to compute $\bU_N$ at the smallest available sampling rate $T_s$. With the proposed framework, we can estimate $\hat{x}_i(T_s)$ for all $i$ where $x_i(T_s)$ is not measured and therefore compute $\bU_N^{T_s}$ and the corresponding $\bL_N$. From the numerical example, we show that multirate DMD approximates the Koopman eigenvalues better than the Koopman eigenvalues computed through the computation of $\bU_N^{MT_s}$, where $M$ is defined as in (\ref{eq:LCM}).
\subsubsection{Numerical Example}
\begin{figure}[]
    \centering
    \includegraphics[scale = 0.45]{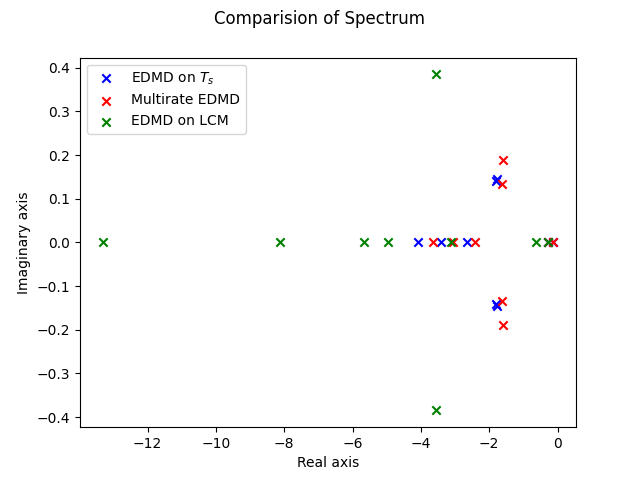}
    \caption{Spectrum of the $\bL_N$ matrix: It can be seen that the spectrum of the infinitesimal generator using multirate EDMD with estimated data better approximates the EDMD spectrum with complete measurements at $x(T_s)$ than the spectrum computed with the data $x(MT_s)$}
    \label{fig1}
\end{figure}

\begin{figure}[]
    \centering
    \includegraphics[scale = 0.5]{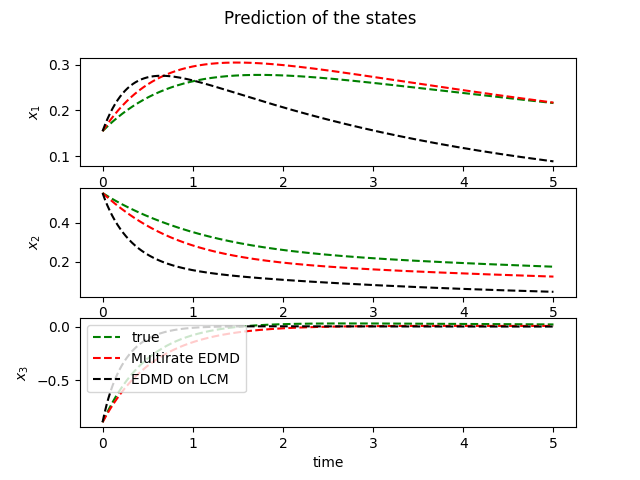}
    \caption{Time prediction with EDMD: The trajectories are better predicted with the proposed multi-rate EDMD than the EDMD performed with data at $x(MT_s)$}
    \label{fig2}
\end{figure}

Consider the following Lorenz system
\begin{align}
 \label{eq:Lorenz} 
 \small{\begin{aligned}
     \dot{x}_1 &= .5(x_2-x_1) \\
     \dot{x}_2 &= x_1(0.75-x_3) - x_2 \\
     \dot{x}_3 &= x_1x_2 - 2x_3.
 \end{aligned}}
\end{align}
\normalsize{}
We assume the sampling rate is $T_s = 0.1$, $x_1$ is measured at all time sampling instants $lT_s$, $x_2$ is measured at sampling instants $4lT_s$ and $x_3$ is measured at sampling instants $3lT_s$, $l = \mathbb{Z}_+$. The data is generated over $300$ sample paths starting from $x_0$ uniformly distributed over $[-1,1]^3$, with $M_2 = 3, M_3 = 4$ (i.e., $3$ and $4$ measurements of the states $x_2$ and $x_3$ in each sample path). Then the estimates $\hat{x}_2(T_s)$ and $\hat{x}_3(T_s)$ at time $T_s$ were computed using Algorithm \ref{alg1}. As $x_1(T_s)$ is measured, we can reconstruct the entire state $\hat{x}(T_s)$ with the estimates $\hat{x}_2(T_s)$ and $\hat{x}_3(T_s)$. Choosing $\{\phi_i\}$ to be monomials up to degree $2$, we performed EDMD and constructed $\bU_N$ and the corresponding $\bL_N$. The spectrum of $\bL_N$ (which are the approximations of the Koopman eigenvalues) are plotted in Figure \ref{fig1}. For comparison, we have also plotted the spectrum of $\bL_N$ under ideal conditions (complete measurements at time $T_s$), and the spectrum of $\bL_N$ corresponding to $\bU_N^{12T_s}$, computed from EDMD performed with data at $x(0)$ and $x(12T_s)$ (as we measure all the states at time instant $12T_s$), denoted as EDMD with LCM. It can seen that, through the proposed multirate EDMD, we are able to obtain better approximations of the spectrum of the infinitesimal generator with complete measurements at $T_s$. Figure \ref{fig2} compares the prediction of trajectories between the proposed multirate EDMD framework with that of the EDMD performed with data measured at the LCM. Moreover multirate EDMD can predict the trajectories more accurately than the EDMD performed at LCM, highlighting the effectiveness of the proposed approach. Note that it was also observed that as the sampling time $T_s$ is increased, the infinitesimal generator corresponding to the restriction of $\bU^{12T_s}$ is complex valued, leading to large inaccuracies in the spectrum of $\bL_N$.

\subsection{Single State Sampling}
Another test case for the proposed EDMD framework with non-uniform sampling is as follows. Given dynamics over $\Rr^n$, at each sampling time $lT_s$, we have access to only one component of the state (and hence the name single state sampling), and each state $x_i$ is measured periodically at time $(i+ln)T_s$, $l = 0,1,\dots,M$. An example of such a measurement in a 3-dimensional system is given in Table \ref{tab2}. 
\begin{table}
       \centering
     \scalebox{0.65}{ \setlength{\extrarowheight}{8pt}%
     \begin{tabular}{||c|c|c|c|c|c|c||}
    \hline
         \backslashbox{States}{Time}&  $T_s$ & $2T_s$ & $3T_s$ & $4T_s$ & $5T_s$ & $6T_s$  \\ \hline 
         $x_1$  & $x_1(T_s)$ & -  &\textcolor{blue}{$\hat{x}_1(3T_s)$}  & $x_1(4T_s)$ & - & - \\ \hline
         $x_2$  & - & $x_2(T_s)$ & \textcolor{blue}{$\hat{x}_2(3T_s)$}& \textcolor{blue}{$\hat{x}_2(4T_s)$} & $x_2(5T_s)$ & -  \\ \hline
         $x_3$  & - & - & $x_3(3T_s)$ & \textcolor{blue}{$\hat{x}_3(4T_s)$}  & - & $x_3(6T_s)$  \\ \hline
    \end{tabular}}
    \caption{Single state sampling: The components in black are measured, and those in blue need to be estimated}
    \label{tab2}
   \end{table}
   
   The main difference between the multi-rate sampling and single state sampling framework is that in the latter, the entire state $x$ is never measured completely at any given sampling instant. Such a scenario can occur in practical systems where only one sensor is used to measure the state components. Hence, at each measurement instant, only one state component is available. In this situation, the traditional EDMD framework can never be performed, and the reconstruction of the state is necessary to perform DMD. For this framework, we see that we have the dead time $r_i = iT_s$ for the state $x_i$, and each state is measured with a sampling rate of $nT_s$. Using the Algorithm \ref{alg1}, we estimate the entire state at time $\hat{x}(nT_s)$ and $\hat{x}((n+1)T_s)$ and perform EDMD with this data.
   

    \subsubsection{Numerical Example}
\begin{figure}[]
    \centering
    \includegraphics[scale = 0.45]{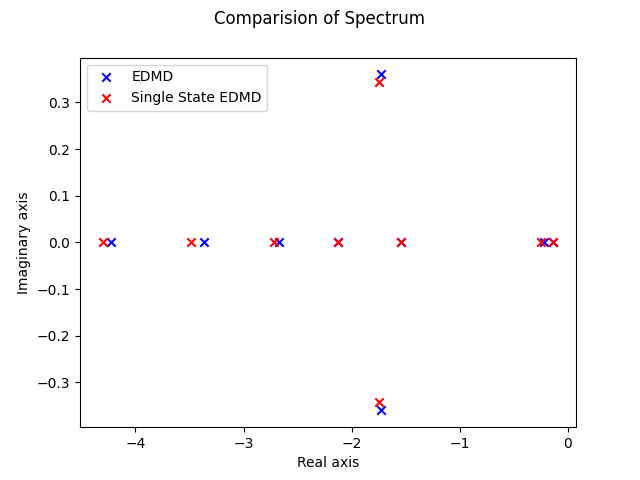}
    \caption{Spectrum of the $\bL_N$ matrix: Comparision of the spectrum of EDMD performed with complete data at $T_s$ and proposed single state EDMD}
    \label{fig3}
\end{figure}

\begin{figure}[]
    \centering
    \includegraphics[scale = 0.5]{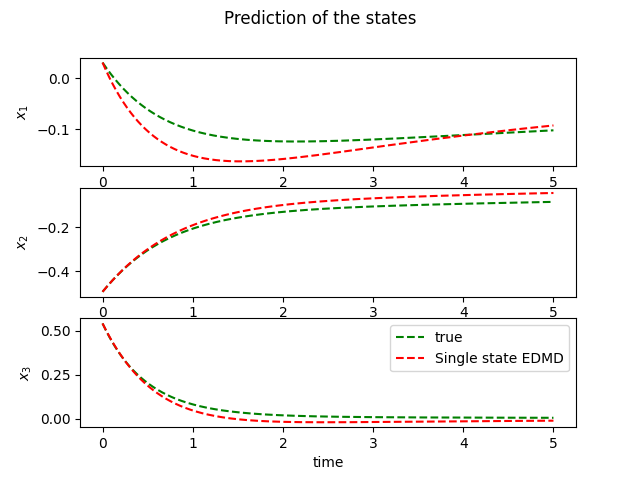}
    \caption{Time prediction with EDMD}
    \label{fig4}
\end{figure}

    In this section, we perform single state EDMD for the system defined in (\ref{eq:Lorenz}). The data is generated at a sampling rate of $T_s = 0.1s$, and each state $x_i$ is measured according to Table \ref{tab2}. $3$ measurements are taken for each state at each of the $100$ sample paths with initial conditions uniformly distributed over $[-1,1]^3$. Using Algorithm \ref{alg1}, we estimate $x_1$ and $x_2$ at time $3T_s$, and estimate $x_2$ and $x_3$ at time $4T_s$. We perform EDMD with the basis functions assumed to be monomials up to degree 2. We compare the results with that of EDMD performed with the complete measurements of states as $\{x^{(k)}(T_s),x^{(k)}(2T_s)\}$. The comparison of the spectrum corresponding to $\bL_N$ is given in Figure \ref{fig3}, and the comparison of time series prediction is given in Figure \ref{fig4}. It can be seen that the proposed single state EDMD can approximate the spectrum of the EDMD performed with complete measurements as well as predict the trajectories of the nonlinear system with a good accuracy. 
    \section{Conclusion}
    In this paper, we discussed an algorithm to compute EDMD and approximate the Koopman operator for a dynamical system when the data is available as partially measured states (and each state component is sampled non-uniformly). This formulation of EDMD is different from the existing DMD techniques for non-uniformly sampled data where the assumption is that, when a data sample is available at a particular time, the entire state is available. The proposed algorithm works in two steps, with the first step aiming to estimate the individual state components at pre-determined times $T_s$ and $2T_s$ from the data through Hankel DMD, hence reconstructing the entire state at $T_s$ and $2T_s$ respectively. The second step performs EDMD with the reconstructed data. Two specific settings of the proposed algorithm are analyzed through numerical simulations for a Lorenz system. In particular, these numerical simulations suggest that our method outperforms a naive application of DMD using a subset of the data, and yields comparable performance to DMD applied on uniformly sampled data. Some important future directions would include extending the algorithm for non-autonomous systems and theoretical analysis of the proposed algorithms.

\end{document}